\documentclass[aps,prb,twocolumn,showpacs]{revtex4}

\usepackage{graphicx}
\usepackage{amsmath}
\usepackage{color}
\usepackage{epstopdf}
\begin{document}

\title{Temperature-dependent and anisotropic optical response of layered Pr$_{0.5}$Ca$_{1.5}$MnO$_{4}$
probed by spectroscopic ellipsometry}

\author{M. A. Majidi$^{1,6}$, E. Thoeng$^{1,2}$, P.K. Gogoi $^{1,2}$, F. Wendt $^{3}$, S. H. Wang $^{1,2}$, I. Santoso $^{1,2}$, T.C. Asmara $^{1,2}$, I.P. Handayani $^{4,5}$, P. H. M. van Loosdrecht$^{4}$, A. A. Nugroho $^{4,5}$, M. R\"{u}bhausen$^{1,3}$, and  A. Rusydi$^{1,2,3,5}$}
\email{phyandri@nus.edu.sg}
\affiliation{$^{1}$ NUSNNI-NanoCore, Department of Physics, Faculty of Science,
National University of Singapore, Singapore 117542, Singapore,}
\affiliation{$^{2}$ Singapore Synchrotron Light Source, National University of Singapore, Singapore 117603,
Singapore}
\affiliation{$^{3}$ Institut f\"{u}r Angewandte Physik, Universit\"{a}t Hamburg, Jungiusstrae 11,
D-20355 Hamburg, Germany. \\
Center for Free Electron Laser Science (CFEL), Notkestra{\ss}e 85,
D- 22607 Hamburg, Germany,}
\affiliation{$^{4}$ Zernike Institute for Advanced Materials, University of Groningen, Nijenborgh 4, 9747 AG Groningen, The Netherlands}
\affiliation{$^{5}$ Physics of Magnetism and Photonics, FMIPA, Institut Teknologi Bandung, Bandung 40132, Indonesia}
\affiliation{$^{6}$ Departemen Fisika, FMIPA, Universitas Indonesia, Depok 16424, Indonesia}
\date{\today}

\begin{abstract}
   We study the temperature dependence as well as anisotropy of optical conductivity ($\sigma_1$) in the pseudocubic single crystal Pr$_{0.5}$Ca$_{1.5}$MnO$_{4}$ using spectrocopic ellipsometry. Three transition temperatures are observed and can be linked to charge-orbital ($T_{\rm CO/OO}$ $\sim$ 320 K), two-dimensional-antiferromagnetic (2D-AFM) ($\sim$ 200 K), and three-dimensional AFM ($T_{\rm N} \sim$ 125 K) orderings. Below $T_{\rm CO/OO}$, $\sigma_1$ shows a 
charge ordering peak ($\sim$0.8 eV) with a significant blue shift as the temperature decreases. 
Calculations based on a model that incorporates a static Jahn-Teller distortion and assumes the existence of a local charge imbalance between two different sublattices support this assignment and explain the blue shift. This view is further supported by the partial spectral weight analysis showing the onset of optical anisotropy at $T_{\rm CO/OO}$ in the charge-ordering region ($0.5-2.5$ eV). 
Interestingly, in the charge-transfer region ($2.5-4$ eV), the spectral weight shows anomalies around the $T_{\rm 2D-AFM}$ that we attribute to the role of oxygen-$p$ orbitals in stabilizing the CE-type magnetic ordering. Our result shows the importance of spin, charge, and lattice degrees of freedom in this layered manganite.
\end{abstract}


\maketitle

\begin{center} \textbf{I. INTRODUCTION} \end{center}

Manganites have attracted intensive research over the past decades due to a wide variety of emerging complex phenomena.\cite{SalamonRMP2001,TokuraRepProg2006} They exhibit a rich phase diagram arising from a strong coupling between charge, orbital, spin, and lattice degrees of freedom.\cite{SalamonRMP2001,TokuraRepProg2006}  Among them, Pr-Ca-Mn-O compounds are of particular interest. The three-dimensional (3D) pseudocubic perovskite Pr$_{1-x}$Ca$_{x}$MnO$_{3}$ undergoes a transition from an antiferromagnetic (AFM) insulator to a ferromagnetic (FM) metal upon hole doping, and also exhibits colossal magnetoresistance. \cite{TomiokaPRB96} On the other hand, in the 2D single-layered system Pr$_{1-x}$Ca$_{1+x}$MnO$_{4}$, the AFM insulating phase is dominant in a wide doping range.\cite{SongxuePNAS07}

One of the most interesting phenomena that can be observed in both systems with half doping is the formation of charge-orbital ordering (CO/OO). Below the CO/OO temperature ($T_{\rm CO/OO}$) charges form a checkerboard pattern with alternating Mn$^{3+}$/Mn$^{4+}$ sites, while the e$_g$ orbitals are aligned in a zig-zag chain.\cite{SongxuePNAS07} Below the Neel temperature ($T_{\rm N}$), the spins of Mn ions are coupled ferromagnetically along the zig-zag chain while they are coupled antiferromagnetically between the chains (CE-type magnetic ordering). In the single-layered manganite Pr$_{0.5}$Ca$_{1.5}$MnO$_{4}$, the CO/OO phase occurs above room temperature at $T_{\rm CO/OO}\sim325$ K,\cite{XZYuPRB2007,Ibarra_solidstatechem,MathieuEPL07} while the 2D-AFM and 3D-AFM phases occur at $T_{\rm 2D-AFM}\sim210$ K and $T_{\rm N}\sim130$ K, respectively. \cite{SongxuePNAS07} 
A neutron study of Chi {\it et al.}\cite{SongxuePNAS07} shows a strong spin-lattice coupling where CO/OO competes with the AFM superexchange interaction. Studies on the optical response related to CE-type phase have also been reported for Pr$_{0.5}$Ca$_{1.5}$MnO$_{4}$\cite{LeePRB2007} and other layered manganites,\cite{IshikawaPRB99, LeePRL2006} in which the authors address the temperature dependence and the anisotropy in the optical conductivity taken from the reflectance measurements up to 3 eV along and perpendicular to the zigzag FM chain.

In this paper, we present our study of the temperature dependence and the anisotropy of the optical conductivity [$\sigma_1(\omega,T)$] and magnetic susceptibility [$\chi(T)$] of Pr$_{0.5}$Ca$_{1.5}$MnO$_{4}$, in a temperature range from above $T_{\rm CO/OO}$ down to below $T_{\rm N}$. 
Note that different from experiments reported in Refs. \onlinecite{LeePRB2007,IshikawaPRB99,LeePRL2006}, here we have used spectroscopic ellypsometry covering an energy range of 0.5 to 4 eV and the measurements were performed along the $a$ and $b$ crystalline axes.
As the temperature decreases, we observe that a peak at 0.8 eV (referred to as a charge-ordering peak) monotonically increases in weight, however its position stays fixed for $T > T_{\rm CO/OO}$ and moves to higher energy (i.e., blue shift) as $T < T_{\rm CO/OO}$. This trend was also observed in other manganites.\cite{LeePRB2007,IshikawaPRB99,LeePRL2006,KunzePRB2003}.
Furthermore, through analysis of partial spectral weights in the charge-ordering ($0.5-2.5$ eV) and the charge-transfer ($2.5-4$ eV) regions, we observe significant changes of the spectral weights in both energy regions when temperature crosses $T_{\rm CO/OO}$, $T_{\rm 2D-AFM}$, and $T_{\rm N}$, indicating changes in the electronic structure of the system occurring prior to each transition. From the anisotropy of the spectral weight changes occurring at those three transition temperatures, we suggest that the charge, orbital, and spin correlations affect the electron effective Mn-O hopping integrals through the deviation of orientation of O-$p$ orbitals with respect to $a$ and $b$ axes.
Although the detailed microscopic mechanism of this interplay has not been fully understood, the idea of a connection between such correlations and the change in electron effective Mn-O hopping integrals is consistent with our earlier studies\cite{RusydiMITPRB08,MajidiPRB11} that explains the temperature dependence of $\sigma_1$ of La$_{0.7}$Ca$_{0.3}$MnO$_3$ up to 22 eV. This outlines the importance of measuring the optical conductivity over a wide energy range or using such an unambiguous method as spectroscopic ellipsometry. \\

\begin{center}\textbf{II. EXPERIMENTS}\end{center}

Single crystals of Pr$_{0.5}$Ca$_{1.5}$MnO$_{4}$ were grown using a floating zone technique. X-ray diffraction measurements were performed to ensure high quality samples and to determine the lattice structure. The magnetic susceptibility $\chi(T)$ was  measured using SQUID-MPMS Quantum Design. The measurement of optical properties was performed with an extended Sentech SE850 ellipsometer. The single crystals Pr$_{0.5}$Ca$_{1.5}$MnO$_{4}$ were mounted in a continuous-He flow cold-finger cryostat operated below a base pressure of 3 x 10$^{-8}$ mbar.\cite{Rauer05} Ellipsometry is a self-normalizing technique 
that measures changes of amplitude and phase of photons being reflected from the sample. 
It provides more accurate and unambiguous results of complex dielectric function without performing Kramers-Kronig transformation.\cite{KunzePRB2003} For spectroscopic ellipsometry measurements, the crystals were cleaved. All measurements were done on two different single crystals with the same composition to ensure the reproducibility.\\

\begin{center}\textbf{II. EXPERIMENTAL RESULTS AND DISCUSSION}\end{center}

\begin{figure}
\includegraphics[width=3.2in]{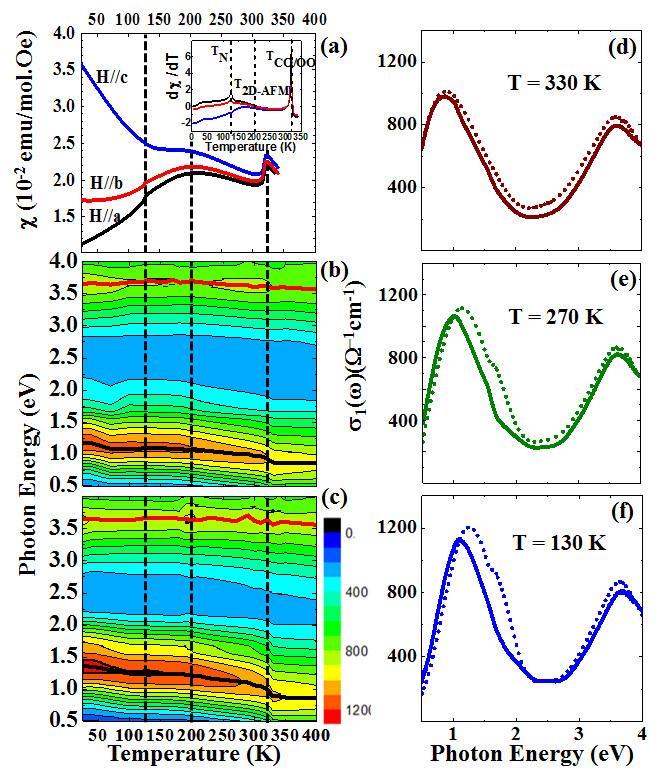}
\caption{(Color online) (a) The temperature dependence of zero-field-cooled magnetic susceptibility ($\chi$) of Pr$_{0.5}$Ca$_{1.5}$MnO$_4$ single crystal along different crystalline axes 
in applied magnetic field of 0.1 T. The inset shows the temperature dependence of the derivative of $\chi$ indicating the magnetic and charge-ordering temperatures. (b) and (c) Contour plots of optical conductivity as a
function of temperature and photon energy, measured along $a$ and $b$ axis, respectively.   
Thick black and red lines connect the positions of charge-ordering and charge-transfer peaks,
respectively, between adjacent temperatures. Dashed black lines indicate three transition temperatures identified from the magnetization measurements. (d)-(f) Anisotropic optical conductivity for $T >T_{\rm CO/OO}$, $T_{\rm CO/OO} > T > T_{\rm 2D-AFM}$, and $T_{\rm 2D-AFM} > T > T_{\rm N}$, 
along $a$ (solid lines) and $b$ (dashed lines) axes.
} 
\label{epsilon contour}
\end{figure}

Fig. \ref{epsilon contour}(a) shows the temperature dependence and anisotropy of $\chi(T)$ of Pr$_{0.5}$Ca$_{1.5}$MnO$_4$ 
along different crystalline axes.
The charge-ordering temperature is clearly observed around 320 K. A broad maximum around 200 K is the signature of short-range two-dimensional AFM correlations.
The long-range AFM ordering is observed at 125 K and the magnetic susceptibility becomes strongly anisotropic below the N\'{e}el temperature. An interesting observation here is the enhancement of anisotropy of $\chi$ for $H||a$ and $H||b$ upon cooling, which is consistent with $\sigma_1$ as discuss below.

Figures \ref{epsilon contour}(b) and (c) give the contour plot of $\sigma_1$ as a function of photon energy and temperature along $a$ and $b$ axis, respectively. They reveal an anisotropic profile of $\sigma_1(\omega)$ as the temperature is varied. Note, that the thick black line connecting the charge-orderingy peak ($\sim0.8$ eV)
between adjacent temperatures clearly 
traces the blue shift that starts as the temperature passes $T_{\rm CO/OO}$ for both measurements along $a$ and $b$ axes. This blue shift signifies a direct connection between the electronic structure revealed in $\sigma_1(\omega)$ and the formation of charge order parameter as outlined later in our calculations.

In order to get a better insight in the anisotropic behavior, in Fig. \ref{epsilon contour} (d)-(f) we plot $\sigma_1(\omega)$ at selected temperatures along both $a$ and $b$ axes. 
Above $T_{\rm CO/OO}$, the anisotropy persists although not so pronounced, indicating that the difference of lattice contants $a$ and $b$ still causes a minor anisotropy. The anisotropic behavior for $T_{\rm CO/OO} > T > T_{\rm 2D-AFM}$ and $T_{\rm 2D-AFM} > T > T_{\rm N}$ looks almost the same in the charge-ordering region ($<2.5$ eV), but their difference is more noticeable in the charge-transfer region (2.5-4 eV). This indicates that photons of charge-transfer energy excite the system in such a way that affect the 2D-AFM ordering. We elaborate more on this along with the discussion related to  Fig. \ref{spectral weight} below.      

Temperature dependent changes in $\sigma_1(\omega)$ for two different energy regions separated by a crossing point at 2.5 eV for both $a$ and $b$ axes can be better shown by the analysis of the partial spectral weight integral ($W$) which describes the effective number of electrons excited by photons of a given energy. The partial spectral weight integral is defined as 
$W = \int_{\omega_1}^{\omega_2}\! \sigma_1(\omega) \, d\omega$, where $\omega_{1}$ and $\omega_{2}$ 
are lower and upper boundaries of the energy region.

\begin{figure}[!t]
\includegraphics[width=3.2in]{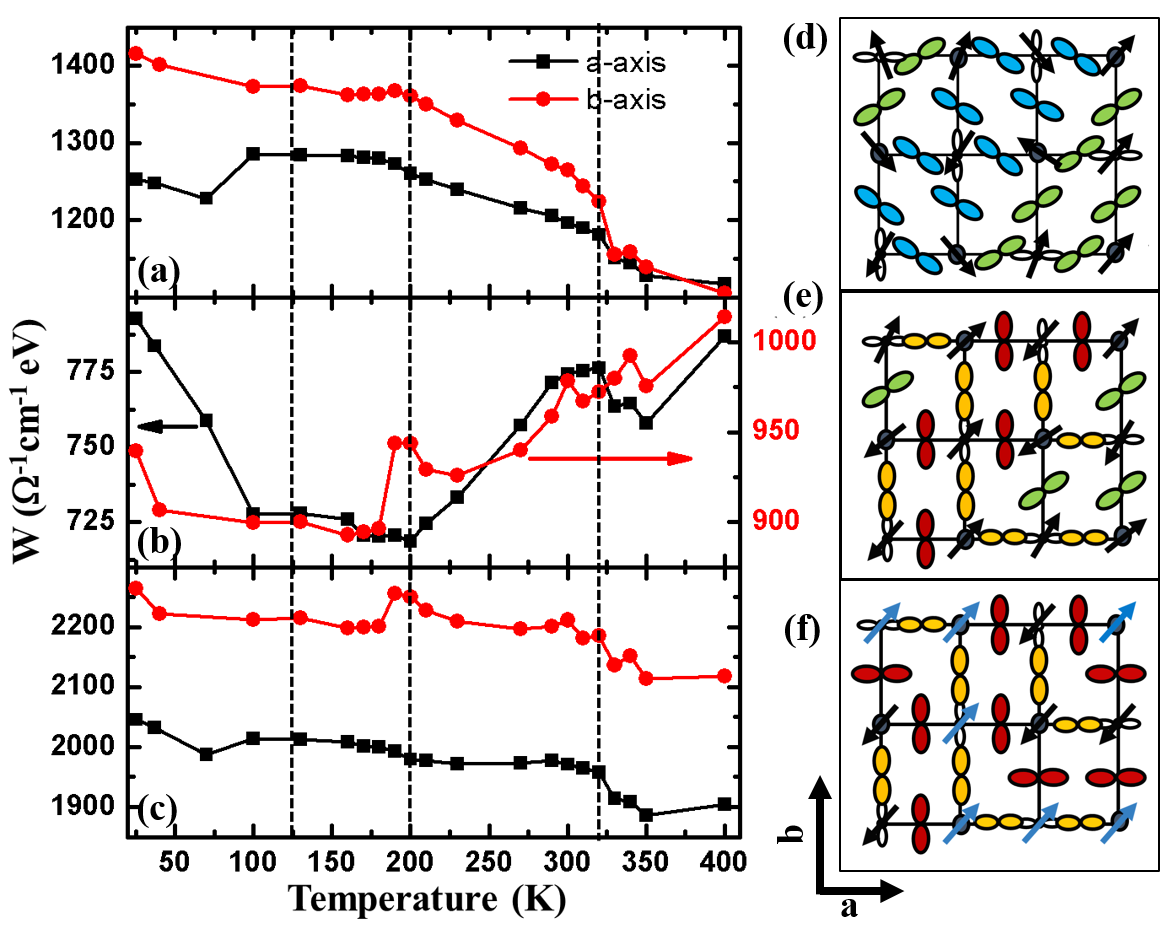}
\caption{(Color online) Temperature dependence of partial spectral weight 
for energy regions: (a) charge ordering (0.5-2.5 eV), (b) charge transfer (2.5-4 eV), and (c) the total integrated spectral weight. Dashed lines indicate the identified transition temperatures in concordance with previous analysis.
Right panels illustrate the process causing the anomalous anisotropy for the charge-transfer region as $T$ crosses $T_{\rm 2D-AFM}\sim 200$K: (d) above, (e) at, and (f) below $T_{\rm 2D-AFM}$. The Mn sites are drawn with arrows depicting their spin directions, while the remainders are the O sites.
}
\label{spectral weight}
\end{figure}

Fig. \ref{spectral weight}(a) shows that the spectral weights in the charge-ordering region, measured along $a$ and $b$ axes practically overlap above $T_{\rm CO/OO}$. As the temperature decreases below $T_{\rm CO/OO}$,
both curves increase abruptly but in an anisotropic fashion. Finally, below $T_{\rm N}$ the two sets of data appear to display more anisotropic behaviors. 

In the charge-transfer region [see Fig. \ref{spectral weight}(b)], both spectral weights decrease 
monotonically from high temperature, behave peculiarly around $T_{\rm CO/OO}$, and decrease further below $T_{\rm CO/OO}$.
Interestingly, the $b$ axis data show remarkable enhancement over the $a$ axis data around $T_{\rm 2D-AFM}$. 
We propose a scenario explaining this rather anomalous behavior in the next paragraph.
With further decrease in temperature, the two sets of data are seen to overlap down to $T_{\rm N}$, before they separate again with a sudden drastic rise of the $a$ axis data. In Fig. \ref{spectral weight}(c), we show that the total integrated spectral weight does not conserve the total number of charge within the measurement energy range. As  $\sigma_1$ is restricted by the $f$-sum rule: $\int_{0}^{\infty} \! \sigma_1(\omega) \, d\omega = \frac{\pi n^{2}}{2m^{*}}$, where $n$ is the number of electrons and $m^{*}$ is the effective mass of electron, our result indicates that the missing charge is to be compensated at much higher energies above 4 eV.

Figures \ref{spectral weight}(d)-(f) illustrate our scenario of the process that yields to the anomalous anisotropy of the partial spectral weight integral for the charge-transfer region as the temperature crosses $T_{\rm 2D-AFM}\sim 200$K. 
At $T$ slightly above $T_{\rm 2D-AFM}$ [see Fig. \ref{spectral weight}(d)], the Mn spins are oriented randomly. 
Thermal and photon-induced excitations distort the orientation of many O-$p$ orbitals from $a$ and $b$ axes, allowing electrons to hop along both axes with roughly equal probability. Hence, the charge-transfer conductivity is roughly equal for both axes.
At $T\sim T_{\rm 2D-AFM}$ [see Fig. \ref{spectral weight}(e)], some of the oxygen orbitals align with $a$ or $b$ axis, so as to effectively mediate the double-exchange (DE) hoppings between the neighboring Mn spins. The system gains energy by ordering the Mn spins in the CE-type AF configuration, howeve, some pairs of O-$p$ orbitals increase their energies associated with $\sigma$-antibonding states that are occupied by electrons participating in the Mn-O charge transfer. To minimize this cost of energy, the majority of O $\sigma$-bonds form along the $b$ axis, rather than $a$ axis (since $b>a$), making the probability of electrons moving straight along the $b$ axis significantly higher than that along the $a$ axis.
Thus, at $T_{\rm 2D-AFM}$ the charge-transfer conductivity along the $b$ axis becomes remarkably higher than that along the $a$ axis, and is the highest among the conductivity values for $a$ and $b$ axes in the three considered temperatures.
Below $T_{\rm 2D-AFM}$ [see Fig. \ref{spectral weight}(f)], the Mn spins are ordered in CE-type AF configuration. To stabilize this Mn spin configuration, the O-$p$ orbitals are oriented in such a way to effectively mediate the DE hoppings between Mn sites with the same spin orientations, and minimize the hopping probability between Mn sites with opposite spin orientations. Here, electrons can only hop along zigzag chains, where only 50\% of O sites contribute to the conductivity.  Thus, the charge-transfer conductivity along $a$ or $b$ axes at this temperature would be roughly equal and is the lowest among those in the three considered temperatures.

On the other hand, we also observe the anisotropic behavior of spectral weight changes below $T_{\rm N}$ in the charge-transfer region. The detailed origin of this behavior is unclear in the present study, however, we argue that the increase of the spectral weight below $T_{\rm N}$ may suggest a stronger Mn-O hybridization for both axes facilitating the stabilization of the 3D-AFM long-range order. Thus, the formation of 2D- and 3D-AFM states requires the interplay of spin, charge, orbital, and lattice degrees of freedom.

\begin{center}\textbf{IV. MODEL AND CALCULATION RESULTS}\end{center}

To address the general profile of the 0-4 eV optical conductivity and the blue shift of the charge-ordering peak, we discuss the dynamics of the electrons of the $e_g$ orbitals and the collective spins of the $t_{2g}$ electrons of the manganese atoms, while the other elements are not considered.
Since the optical response comes mainly from the MnO$_2$ plane,
we model the system as a two-dimensional square lattice divided into sub-lattices $A$ and $B$ to accommodate the formation of charge ordering.
Each sub-lattice atom has two $e_g$
orbitals, and each orbital has two possible spin orientations, hence
we choose eight basis orbitals to construct the Hilbert space, which we order in
the following way:
$|{\rm Mn}^A ~ e_{g~ 1,\uparrow}\rangle$,
$|{\rm Mn}^B ~ e_{g~ 1,\uparrow}\rangle$,
$|{\rm Mn}^A ~ e_{g~ 2,\uparrow}\rangle$,
$|{\rm Mn}^B ~ e_{g~ 2,\uparrow}\rangle$,
$|{\rm Mn}^A ~ e_{g~ 1,\downarrow}\rangle$,
$|{\rm Mn}^B ~ e_{g~ 1,\downarrow}\rangle$,
$|{\rm Mn}^A ~ e_{g~ 2,\downarrow}\rangle$, and
$|{\rm Mn}^B ~ e_{g~ 2,\downarrow}\rangle$.
Here, $|~ e_{g~ 1}\rangle$ and $|~ e_{g~ 2}\rangle$ are the original $e_g$ states having degenerate on-site energies, $E_0$ (set to 0 eV), in the absence of Jahn-Teller distortion. The effect of the assumed static Jahn-Teller distortion is incorporated 
in terms of the displacements $u$ and $v$ corresponding to two vibration modes as also used in Ref. \onlinecite{LeePRL2006}. Here, we parameterize the static equilibrium values of $u$ and $v$ with $U_{\rm JT}$ and $V_{\rm JT}$, respectively.
Further, we take into account the on-site Coulomb interactions between electrons in the two orbitals with a Hubbard parameter $U$. 
We also consider the on-site Coulomb interactions between electrons in the same orbital with a Hubbard
parameter $U_{1(2)}$ for $|~ e_{g~ 1(2)}\rangle$ orbital. Upon taking $U_1$ and $U_2$ to be infinity, the double occupancy is prevented in each of the orbitals 1 and 2, thus making the total effective number of states accessible to electrons per unit supercell A-B equal to 4, instead of 8. 
To simplify the calculations, we treat the Jahn-Teller and the on-site inter-orbital Hubbard $U$ terms within the Hartree-Fock approximation.
Using the aforementioned basis set we propose the following effective Hamiltonian: 
\begin{eqnarray}
\label{eq:H_eff}
H=\frac{1}{N}\sum_{\bf k} \eta^\dagger_{\bf k} [H_0({\bf k})]_{\rm eff}\eta_{\bf k} - \sum_{i,s,\alpha} J {\bf S}_i^{s}{\bf .s}_{i \alpha}^{s} .
\end{eqnarray}

The first term in Eq. (\ref{eq:H_eff}) is the effective kinetic part, whereof $\eta^\dagger_{\bf k}$ ($\eta_{\bf k}$) is a row (column) vector whose elements are the creation (annihilation) operators associated with the eight basis orbitals. $[H_0({\bf k})]_{\rm eff}$ is an 8$\times$8 matrix in momentum space whose structure
is arranged in four 4$\times$4 blocks
corresponding to their spin directions as
\begin{eqnarray}
\label{eq:H0_eff}
 [H_0({\bf k})]_{\rm eff}=
\left[
\begin{matrix}
H_{\rm HF}({\bf k})_\uparrow&{\bf O}\\
{\bf O} &H_{\rm HF}({\bf k})_\downarrow
\end{matrix}
\right],
\end{eqnarray}
where ${\bf O}$ is a zero matrix of size 4$\times$4, and
\begin{widetext}
\begin{small}
\begin{eqnarray}
\label{eq:HHF_4x4}
H_{\rm HF}({\bf k})_{\uparrow(\downarrow)} =
\left[
\begin{matrix}
E_0+\langle n_1^A\rangle U_{\rm JT}+\langle n_2^A\rangle U & \epsilon_{11}(k_x,k_y) & V_{\rm JT} &\epsilon_{12}(k_x,k_y)\\
\epsilon_{11}(k_x,k_y) & E_0-\langle n_1^B\rangle U_{\rm JT}+\langle n_2^B\rangle U & \epsilon_{12}(k_x,k_y) & V_{\rm JT} \\
V_{\rm JT} & \epsilon_{12}(k_x,k_y) & E_0+\langle n_2^A\rangle U_{\rm JT}+\langle n_1^A\rangle U & \epsilon_{22}(k_x,k_y) \\
\epsilon_{12}(k_x,k_y) & V_{\rm JT} & \epsilon_{22}(k_x,k_y) & E_0-\langle n_2^B\rangle U_{\rm JT}+\langle n_1^B\rangle U
\end{matrix}
\right].
\end{eqnarray}
\end{small}
\end{widetext}
The ${\bf k}$ dependence of $H_{\rm HF}({\bf k})_{\uparrow(\downarrow)}$ comes from the in-plane tight-binding energy dispersions
\begin{eqnarray}
\epsilon_{\alpha \beta}(k_x,k_y)=-2t_{\alpha \beta}(\cos2k_x+\cos2k_y),
\end{eqnarray}
where $t_{\alpha \beta}$, with $\alpha,\beta\in \{1,2\}$, are the effective hopping integrals 
connecting orbital $\alpha$ at a site and orbital $\beta$ at its nearest neighbor site, or vice versa.

The Pr$_{0.5}$Ca$_{1.5}$MnO$_{4}$ system has one $e_g$ electron per unit supercell A-B, hence the electron filling satisfies
$\langle n_1^A \rangle +\langle n_2^A \rangle +\langle n_1^B \rangle +\langle n_2^B\rangle = 1$.
Further, we impose a set of constaints
$\langle n_1^A \rangle =0$, $\langle n_2^B \rangle =0$,
$\langle n_2^A\rangle =\frac{1+\Delta n}{2}$, and  $\langle n_1^B \rangle =\frac{1-\Delta n}{2}$,
with $\Delta n = \langle n_2^A\rangle -\langle n_1^B\rangle $ being a local charge imbalance between two different sublattices.
As this local quantity repeats throughout the entire crystal, its existence is a precondition to a charge ordering.
We assume $\Delta n$ to have a "mean-field" temperature dependence of the form
$(1-\frac{T}{T_{\rm CO/OO}})^{1/2}$.
Note, the model accommodates a charge ordering (CO), but not orbital ordering (OO).
We use the symbol $T_{\rm CO/OO}$ in our calculations just so we can compare its role with that in the experimental data.

Finally, the second term  in Eq. (\ref{eq:H_eff}) represents the exchange interactions between the local spin ${\bf S}_i^s$ of Mn of sub-lattice $s$ 
at a supercell $i$, formed by the strong Hund's coupling among three $t_{2g}$ electrons giving {\it S}=3/2, 
and the electron spin ${\bf s}_{\alpha i}^{s}$ occupying orbital $\alpha$. These interactions are treated within the dynamical mean-field theory
\cite{DMFTGeorgesRMP96}$^,$\cite{Furukawa94}, restricted to the paramagnetic phase.

\begin{figure}
\includegraphics[width=3.0in]{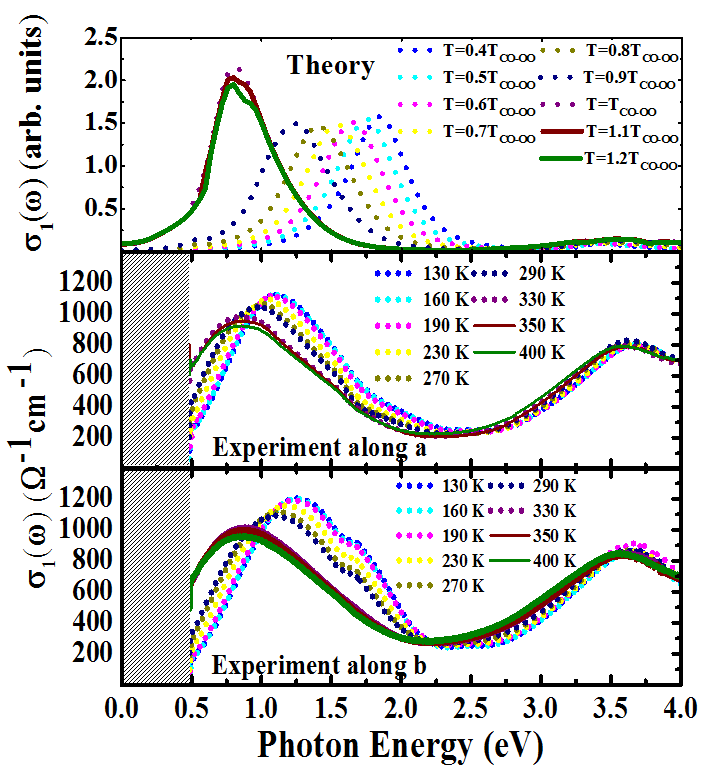}
\caption{(Color online) Temperature dependence of optical conductivity.
The top panel shows the calculation results, while the middle and bottom
panels show the experimental results obtained along $a$ and $b$ axis, respectively. 
}
\label{T-dep}
\end{figure}

We calculate $\sigma_1(\omega)$ for various temperatures using the Kubo formula as done in Ref. \onlinecite{MajidiPRB11},
with the parameter values of $E_0=0$ eV, $t_{11}=t_{22}=0.5$ eV, $t_{12}=0.25$ eV, $U_{\rm JT}=1.2$ eV, $V_{\rm JT}=2$ eV, $U=4$ eV, and $J=0.6$ eV.
The results are shown in Fig. \ref{T-dep}.

As shown in Fig. \ref{T-dep}, our calculations capture the general profile of $\sigma_1(\omega)$, 
most notably the appearance of a charge-ordering peak at 0.8 eV and a charge-transfer peak at 3.6 eV, 
and the blue shift of the charge-ordering peak as the temperature decreases below $T_{\rm CO/OO}$. 
According to our model, the low-energy peak arises along with the formation of Jahn-Teller gap, while its blue shift 
results from the formation of local charge imbalance. 
This blue shift and the charge ordering itself are only possible in the presence of the on-site interorbital Hubbard $U$,
in consistency with the picture described in Ref. \onlinecite{BrinkPRL99}.
Within our model, the charge-transfer peak at 3.6 eV corresponds to transitions from singly to doubly occupied states in Mn sites.
Since this peak appears more pronounced in the experimental data, there may be other contributions 
to this peak involving O bands that we negtect.
Our calculatons also capture the increasing trend in the intensity of the charge-ordering peak as the temperature decreases, with a little discrepancy that an abrupt decrease occurs as the temperature is lowered just slightly below $T_{\rm CO/OO}$. This may also be a consequence of the neglect of the role of O orbitals.
The middle and bottom panels of Fig. \ref{T-dep} show the anisotropy along $a$ 
and $b$ axes, which our model does not capture. \\

\begin{center} \textbf{V. CONCLUSION} \end{center}

In conclusion, we have observed the temperature dependence and the anisotropy of $\sigma_1(\omega)$ and $\chi$ of Pr$_{0.5}$Ca$_{1.5}$MnO$_{4}$ measured along different crystalline axes. 
The anisotropic behavior of partial spectral weight of charge-ordering region 
of $\sigma_1(\omega)$ as a function of temperature is similar to that of $\chi$, indicating the strong
connection between the optical response and the charge and spin correlations.
The observation of a blue shift of the charge-ordering peak of $\sigma_1(\omega)$ below 
$T_{C\rm O/OO}$ signifies its connection with the formation of a charge-order parameter, which is in agreement with our calculations. Analysis of the prominent anisotropic change of partial spectral weight around $T_{\rm 2D-AFM}$ suggests an interplay between spin, charge, and orbital correlations through the change of the electron effective Mn-O hopping integrals along the $a$ and $b$ axes. \\

\begin{center} \textbf{ACKNOWLEDGEMENT} \end{center}

We acknowledge Tjia May On for the discussions and his valuable comments. This work is supported by Singapore National Research Foundation under its Competitive Research Funding (NRF-CRP 8-2011-06 and NRF2008NRF-CRP002024), MOE-AcRF-Tier-2, NUS-YIA, FRC, BMBF under 50KS7GUD as well as DFG through Ru 773/5-1. We acknowledge the CSE-NUS computing centre for providing facilities for our numerical calculations.

\end{document}